\documentclass[11pt]{article}
\pdfoutput=1
\usepackage{jheppub}
\usepackage{amssymb,amsfonts}
\usepackage{mathrsfs}
\usepackage{epsfig}
\let\savenumberline\numberline
\def\numberline#1{\savenumberline{#1.}}
\makeatletter
\renewcommand{\@seccntformat}[1]{\csname the#1\endcsname.\,\,}
\makeatother

\newcommand{\CB}{{\cal B}}
\newcommand{\CD}{{\cal D}}
\newcommand{\CE}{{\cal E}}
\newcommand{\CF}{{\cal F}}
\newcommand{\CG}{{\cal G}}

\newcommand{\CL}{{\cal L}}
\newcommand{\CM}{{\cal M}}
\newcommand{\CN}{{\cal N}}

\newcommand{\CW}{{\cal W}}

\newcommand{\RD}{{\mathrm D}}
\newcommand{\barRD}{{\bar{\mathrm D}}}
\newcommand{\SB}{{\mathscr B}}
\newcommand{\SD}{{\mathscr D}}
\newcommand{\SE}{{\mathscr E}}
\newcommand{\SF}{{\mathscr F}}
\newcommand{\SG}{{\mathscr G}}
\newcommand{\SM}{{\mathscr M}}
\newcommand{\ST}{{\mathscr T}}

\newcommand{\p}{\partial}
\renewcommand{\bar}[1]{\overline{#1}}
\renewcommand{\tilde}[1]{\widetilde{#1}}
\newcommand{\be}{\begin{equation}}
\newcommand{\ee}{\end{equation}}
\newcommand{\bea}{\begin{eqnarray}}
\newcommand{\eea}{\end{eqnarray}}

\newcommand{\ie}{{\it i.e.}}

\newcommand{\diff}{{\rm Diff}}

\makeatletter
\def\@fpheader{\relax}
\makeatother
\usepackage{graphicx}
\usepackage{latexsym}

\title{\ \vspace{1.5in} \\ Perelman's Ricci Flow in Topological Quantum Gravity}
\author{Alexander Frenkel${}^a$, Petr Ho\v{r}ava${}^{b,c}$ and Stephen Randall${}^{b,c}$}
\affiliation{${}^a$Stanford Institute for Theoretical Physics and Department of Physics\\
Stanford University, Stanford, CA, 94305-4060, USA\medskip\\
${}^b$Berkeley Center for Theoretical Physics and Department of Physics\\
University of California, Berkeley, CA, 94720-7300, USA\medskip\\
${}^c$Physics Division, Lawrence Berkeley National Laboratory\\
Berkeley, CA 94720-8162, USA}
\abstract{We find the regime of our recently constructed topological nonrelativistic quantum gravity, in which Perelman's Ricci flow equations on Riemannian manifolds appear precisely as the localization equations in the path integral.  In this mapping between physics and mathematics, the role of Perelman's dilaton is played by our lapse function.  Perelman's local fixed volume condition emerges dynamically as the $\lambda$ parameter in our kinetic term approaches $\lambda\to-\infty$.  The DeTurck trick that decouples the metric flow from the dilaton flow is simply a gauge-fixing condition for the gauge symmetry of spatial diffeomorphisms.  We show how Perelman's ${\cal F}$ and ${\cal W}$ entropy functionals are related to our superpotential.  We explain the origin of Perelman's $\tau$ function, which appears in the ${\cal W}$ entropy functional for shrinking solitons, as the Goldstone mode associated with time translations and spatial rescalings: In fact, in our quantum gravity, Perelman's $\tau$ turns out to play the role of a dilaton for anisotropic scale transformations.  The map between Perelman's flow and the localization equations in our topological quantum gravity requires an interesting redefinition of fields, which includes a reframing of the metric.  With this embedding of Perelman's equations into topological quantum gravity, a wealth of mathematical results on the Ricci flow can now be imported into physics and reformulated in the language of quantum field theory.}
\begin{document}
\maketitle
\section{Introduction}

This paper is a sequel to our previous work \cite{grf}, in which we presented a topological nonrelativistic quantum gravity associated with the general theory of Ricci flow on Riemannian manifolds.  

This topological gravity is of the cohomological type \cite{ewcoho}, with the action and the path integral generated entirely in the process of BRST gauge fixing of the underlying gauge symmetries, and whose original Lagrangian is zero (or, at most, a sum of topological invariants).  Besides the existence of the BRST charge $Q$, in \cite{grf} we also required the existence of an extended BRST algebra, with $\CN=2$ supercharges $Q$ and $\bar Q$ and the anticommutation relations
\be
Q^2=0,\qquad \bar Q^2=0,\qquad \{Q,\bar Q\}=\p_t.
\label{ee2susy}
\ee
The appearance of the generator of time translations in the anticommutator of $Q$ and $\bar Q$ is consistent with the fact that our theory is nonrelativistic, and the underlying spacetime manifold $\CM$, of dimension $D+1$, carries a preferred foliation $\SF$ by constant time slices of dimension $D$.  For now, we consider only those spacetime geometries whose leaves of constant time are all isomorphic to a given compact $D$-dimensional spatial manifold $\Sigma$, leaving the questions of topology-changing transitions or of boundary conditions for noncompact $\Sigma$ to future work.  In the simplest version of this topological gravity, the dynamical field is the spatial metric $g_{ij}$, but it is much more interesting to extend the theory to a gauge theory with foliation-preserving spacetime diffeomorphisms, and the dynamical fields corresponding to the full spacetime metric, decomposed into $g_{ij}$, the shift vector $n^i$, and the lapse function $n$.  These will be the dynamical variables in our quantum gravity throughout the rest of the paper. 

The postulate of $\CN=2$ BRST supersymmetry makes it natural to formulate the theory in $\CN=2$ superspace, and that was the strategy we followed in \cite{grf}.  The superpartners of the basic fields in this superspace are the various ghosts, antighosts and auxiliary fields known from the BRST formalism.  We further clarified the structure of the component fields, the underlying gauge symmetries, and the BRST gauge fixing process in \cite{gst}.

Having constructed all this machinery of topological quantum gravity theory, it remains to show how the various celebrated mathematical examples of Ricci flow equations on Riemannian manifolds -- especially Perelman's Ricci flow -- emerge in this quantum theory.  This is the task of the present paper.

\subsection{Ricci flows on Riemannian manifolds}

Since its inception in the 1980's \cite{hamfirst}, the mathematical theory of the Ricci flow on Riemannian manifolds has undergone several stages of development.  The first stage, lasting for about two decades, was dominated by the study of Hamilton's Ricci flow equation,
\be
\frac{\p\hat g_{ij}}{\p t}=-2\hat R_{ij},
\ee
and its geometrical consequences.  Many important achievements highlight this era \cite{collrf}.  The next stage was reached with Perelman's Ricci flow equations \cite{perel1,perel2,perel3} (see \cite{hrf,rfi,rf1,rf2,rf3,rf4,topping,muller} for extensive reviews), which are designed so that the flow of the metric is coupled to the flow of another field, which Perelman called the ``dilaton.''  (This field is traditionally denoted by $f$, but we will call it $\hat\phi$ in this paper.)%
\footnote{Throughout this paper, we will systematically denote Perelman's variables by hats, $\hat{\ }$. This includes both the geometric fields $\hat g_{ij}$, $\hat\phi$, $\ldots$ and the various geometric quantities such as the covariant derivative $\hat\nabla_i$.  We reserve the notation without hats for our variables $g_{ij}$, $\phi$, $\ldots$, $\nabla_i$.  These two sets of variables will be related by a nonlinear transformation which involves a change of frame for the metric.}
\bea
\label{eepone}
\frac{\p\hat g_{ij}}{\p t}&=&-2\hat R_{ij}-2\hat\nabla_i\p_j\hat\phi,\\
\label{eeptwo}
\frac{\p\hat\phi}{\p t}&=&-\hat R-\hat\Delta\hat\phi.
\eea
The major advance in Perelman's formulation stems from the fact that the right-hand side of these coupled flow equations is given by the gradient of a functional, Perelman's ``$\CF$-functional''
\be
\hat\CF=2\int d^D x\sqrt{\hat g}e^{-\hat\phi}\left\{\hat R+\hat g^{ij}\p_i\hat\phi\p_j\hat\phi\right\},
\label{eeperelf}
\ee
assuming that the variations of $\hat g_{ij}$ and $\hat\phi$ are subjected to the constraint requiring that the volume element
\be
\label{eefvol}
e^{-\hat\phi}\sqrt{\hat g}\,d^D x=dm(x^i)
\ee
be held fixed in time and equal to a fixed measure $dm(x^i)$ on the spatial manifold $\Sigma$.

Perelman's equations are further simplified by an application of what has become known in the mathematical literature as ``DeTurck's trick'':  a specific spatial diffeomorphism is applied to the original equations, with its generating vector field $\xi^i$ given by the gradient of the dilaton, $\xi^i=\hat g^{ij}\p_j\hat\phi$.  After this diffeomorphism, Perelman's Ricci flow equations simplify to
\bea
\frac{\p\hat g_{ij}}{\p t}&=&-2\hat R_{ij},\label{eepdtone}\\
\frac{\p\hat\phi}{\p t}&=&-\hat R-\hat\Delta\hat\phi+\hat g^{ij}\p_i\hat\phi\p_j\hat\phi,\label{eepdttwo}
\eea
In this form, Hamilton's original metric flow equation is now nicely separated from the flow equation for the dilaton.

Perelman's Ricci flow was of course instrumental in his proof of the Poincar\'e conjecture \cite{perel3,morgant1,morgantb,morgant2,morgant3}, and since then in the proofs of many other important results: the Thurston geometrization conjecture \cite{perel3,morgantb}, the generalized Smale conjecture \cite{smale,bamkl1,bamkl2,kl1,kl2}, and even a new proof of the uniformization theorem in two spatial dimensions \cite{tiann}.  

\subsection{Topological nonrelativistic gravity and the Ricci flow}

The topological quantum gravity theory presented in \cite{grf} has been designed around a family of generalized Ricci flow equations similar to Perelman's.  These equations appear in the topological gravity in a central role, as localization equations:  With the appropriate initial or boundary conditions, the path integral is reduced by standard topological arguments to an integral over the space of classical solutions to the flow equations.

The localization equations depend on many physical coupling constants available in this theory.  One can naturally ask whether Perelman's equations can be precisely reproduced in some regime of this topological quantum gravity, and if so, what is the precise mapping of the variables between the mathematical and the physical picture.  In the present paper, we address these questions in the semiclassical limit of the theory.

That such a direct embedding of Perelman's Ricci flow equations into our topological gravity should even exist is not immediately obvious, for several reasons.  First, note that Perelman's equations have less spacetime symmetry than the localization equations of the theory constructed in \cite{grf}, at least before we fix a part of the secondary gauge symmetry of foliation-preserving spacetime diffeomorphisms.  At best, we can find a covariantized version of Perelman's equations which respects time-reparametrization invariance (this will be done in Section~\ref{seccpf}); or we need to propose an appropriate gauge fixing of time reparametrizations in our theory to match the symmetries of Perelman's equations (this will be the subject of Section~\ref{secpf}).  Secondly, our version of the flow equations is schematically of the form
\be
\label{eesch}
e^\phi\dot g_{ij}=-\alpha_RR_{ij}+\ldots,
\ee
with an extra multiplicative $e^\phi$ factor between the two sides.  This suggests the need for a nonlinear ``reframing'' field redefinition between Perelman's variables and ours.  Thirdly, the question is how to interpret the additional volume-fixing condition (\ref{eefvol}), which was postulated by Perelman in order to derive the flow equations.  This condition cannot be a part of gauge fixing of the residual gauge symmetries in our theory, since (i) the theory presented in \cite{grf} only exhibits \textit{spatially-independent} time reparametrizations, and (ii) we wish to reserve all the spatial diffeomorphism symmetry in the unfixed form, so that we can implement the DeTurck trick as a gauge fixing condition, as we did in \cite{grf}.  Finally, note also that in our approach, \textit{both} sides of the flow equations come from a variational principle, which leads not only to more couplings but also to additional restrictions on the form of the equations of motion one can obtain.

In this paper, we resolve these issues in two steps.  First, in Section~\ref{seccpf}, we consider the covariant theory of \cite{grf}, with secondary gauge symmetry of foliation-preserving spacetime diffeomorphisms.  We identify the regime where the localization equations represent the covariant version of Perelman's equations.  In the process, we learn how our fields are related to Perelman's by a change-of-frame transformation, and how Perelman's fixed-volume condition emerges dynamically in our topological gravity.  Then, in Section~\ref{secpf}, we perform the gauge fixing that leads directly to the localization of the path integral on Perelman's Ricci flow equations (\ref{eepone}) and (\ref{eeptwo}), and the fixed-volume condition (\ref{eefvol}).  Those readers who are interested only in our final product -- the topological gravity of Perelman's Ricci flow -- can go directly to Section~\ref{secpf}, and return to Section~\ref{seccpf} for the motivation and logical derivations as needed.  In Section~\ref{sectime} we extend our construction to include the $\CW$ and $\CW_+$ entropy functionals associated with the shrinking and expanding Ricci solitons; the main new ingredient that we will need to introduce is going to be the Goldstone superfield $T$ asssociated with spontaneously broken time translations.  

\section{Covariantized Perelman-Ricci flow equations from topological gravity}
\label{seccpf}

The theory presented in \cite{grf} and further studied in \cite{gst} is a theory of the spacetime metric expressed in the ADM variables \cite{adm}, consisting of the spatial metric $g_{ij}$, the shift vector $n^i$, and the lapse function $n$.  The theory is designed to be topologically invariant, and nonrelativistic -- it is sensitive to a preferred foliation $\SF$ of spacetime $\CM$ by leaves $\Sigma$ of constant time.  Given the required $\CN=2$ extension (\ref{ee2susy}) of the BRST symmetry, this topological theory is most concisely formulated in an $\CN=2$ superspace extension \cite{grf} of the nonrelativistic spacetime manifold $\CM$.  Spacetime is thus extended to a supermanifold $\SM$ of superdimension $(D+1|2)$, which also inherits a foliation by the spatial leaves $\Sigma$.

Throughout this section, we will consider the theory which enjoys -- besides the topological symmetry -- a secondary gauge symmetry of foliation-preserving diffeomorphisms of spacetime, 
\be
\CG\equiv \diff_\SF(\CM).
\label{eedifff}
\ee
This symmetry is locally generated by infinitesimal spacetime-dependent spatial diffeomorphisms $\delta x^i=\xi^i(t, x^j)$, and time-dependent time reparametrizations $\delta t=f(t)$.  We intend to treat this secondary symmetry equivariantly:  In particular, we will construct our action functionals in this section to be manifestly invariant under $\CG$.  

As is often done in supersymmetric gauge-theory constructions, we impose this secondary gauge symmetry $\CG$ by first extending it to a gauge symmetry in superspace, making $\CG$ manifestly consistent with the underlying $\CN=2$ supersymmetry.  The details of the construction can be found in \cite{grf}, and we do not repeat them here; also, unless stated otherwise, we use the same notation as in \cite{grf}. The full list of superfields in this theory consists of the unconstrained metric superfield $G_{ij}$, whose lowest component is $g_{ij}$, the superfields $N^i$, $S^i$ and $\bar S^i$ in the shift-vector sector, with the bosonic shift vector $n^i$ appearing as the lowest component of $N^i$; and the superfields $E$, $\Theta$ and $\bar\Theta$ in the lapse function sector, with the lapse function $n$ appearing as the inverse of the lowest component $e$ of $E$, $n=1/e$.  In components, each of the three sectors contains the original ADM bosonic field, its ghost, its antighost, and an auxiliary.  In the spatial metric sector, these component fields are
\be
g_{ij},\quad\psi_{ij},\quad\chi_{ij},\quad B_{ij}.
\label{eegcomp}
\ee
In the lapse and shift sectors, they are
\be
\label{eencomp}
e\equiv 1/n,\quad \nu,\quad\bar\nu,\quad w
\ee
and
\be
\label{eenicomp}
n^i,\quad\sigma^i,\quad\bar\sigma^i,\quad X^i,
\ee
with $B_{ij}$, $w$ and $X^i$ the bosonic auxiliaries in the corresponding sectors.  
See \cite{grf,gst} for all additional details, including the $Q$ and $\bar Q$ transformation rules for these component multiplets, and for our conventions.  

\subsection{Localization equations for $g_{ij}$ in topological quantum gravity}

The action can be written as \cite{grf}
\be
S=\frac{1}{\kappa^2}\left(S_K-S_\CW\right),
\ee
where the kinetic term $S_K$ is defined as the part of the action containing at least one supertime derivative, and the term $S_\CW$ -- which we call the superpotential -- contains all the remaining terms, with no supertime derivatives.  The minimal kinetic term is
\be
\label{eecovsk}
S_K=\int d^2\theta\,dt\,d^Dx\,\sqrt{G}N\,(G^{ik}G^{j\ell}-\lambda G^{ij}G^{k\ell})\,\SD_{\bar\theta}G_{ij}\,\SD_\theta G_{k\ell}.
\ee
Here $\SD_{\bar\theta}$ and $\SD_\theta$ are the covariant superspace derivatives associated with the gauge group $\CG$ \cite{grf}.

The superpotential terms can be organized by the increasing dimension of the operators.  In the case of interest, relevant to Ricci flow, we focus on all the terms up to second order in spatial derivatives, which gives -- up to integration by parts -- the following superpotential:
\be
\label{eesuperp}
S_\CW=\int d^2\theta\,dt\,d^Dx\,\sqrt{G}N\left\{\alpha_RR^{(G)}+\alpha_\Phi G^{ij}\p_i\Phi\p_j\Phi+\alpha_\Lambda\right\},
\ee
where $\alpha_R,\alpha_\Phi$ and $\alpha_\Lambda$ are real coupling constants, and the superfield $R^{(G)}$ is the Ricci scalar superfield constructed from the metric superfield $G_{ij}$.  The supefield $\Phi$ is simply related to the lapse superfield $N\equiv 1/E$,
\be
\Phi\equiv -\log N\equiv\log E.
\ee
Since we have restricted our attention to the terms in $S_\CW$ with up to two spatial derivatives, we are anticipating that the short-distance behavior in this theory will exhibit anisotropy between time and space characterized by the dynamical exponent $z=2$.  With this scaling, the two-derivative terms in $S_\CW$ are of the same classical scaling dimension as the kinetic term.  The cosmological constant term $\alpha_\Lambda$ is the only available relevant term.  

Now we wish to study the localization equations, and their possible relation to Perelman's Ricci flow.  We begin with a simple theory, in which the lapse superfield $E$ is constrained to be chiral, and no chirality conditions are imposed on either $G_{ij}$ or $N^i$; extending the terminology of \cite{grf}, we shall refer to this theory as Type C (for ``chiral'' lapse).  First, we will focus on the localization equation for the spatial metric $g_{ij}$ in this Type C theory.  In terms of the component fields listed in (\ref{eencomp}), the chirality condition amounts to setting the antighost $\bar\sigma^i$ and the auxiliary $w$ to zero.

When focusing on the localization equations, we will often -- for clarity of discussion -- keep track of only the bosonic fields, setting all the fermionic component fields to zero.%
\footnote{We use the symbol ``$\approx$'' to denote the evaluation of any quantity by keeping its full dependence on the bosonic component fields while setting all the fermionic components to zero.} 
When we write the type C theory in components, it will be consistent with our equivariant treatment of spatial diffeomorphisms to adopt Wess-Zumino gauge in the superspace shift sector, as discussed in \cite{grf}.  This choice is equivalent to setting $\sigma^i$, $\bar\sigma^i$ and $X^i$ in (\ref{eenicomp}) to zero.  In this gauge, the bosonic gauge symmetry $\CG$ of the action is still manifest.  In particular, the time derivative of $g_{ij}$ is still covariantized to
\be
\nabla_t g_{ij}\equiv\p_t g_{ij}-\nabla_in_j-\nabla_jn_i.
\ee
With this notation, the bosonic part of the action is
\be
S_K\approx-\int dt\,d^Dx\,\sqrt{g}n\,(g^{ik}g^{j\ell}-\lambda g^{ij}g^{k\ell})\,B_{ij}B_{k\ell}
+\int dt\,d^Dx\,\sqrt{g}\,(g^{ik}g^{j\ell}-\lambda g^{ij}g^{k\ell})\,B_{ij}\nabla_tg_{k\ell}
\ee
and
\be
S_\CW\approx\int dt\,d^Dx\sqrt{g}n\,B_{ij}\CE^{ij},
\ee
with $\CE^{ij}$ given by 
\bea
\CE^{ij}\equiv\frac{1}{\sqrt{g}n}\frac{\delta\CF}{\delta g_{ij}}&=&\alpha_R\left(\frac{1}{2}Rg^{ij}-R^{ij}\right)+\alpha_R\left(g^{ik}g^{j\ell}-g^{ij}g^{k\ell}\right)\frac{1}{n}\nabla_k\p_\ell n\nonumber\\
&&\qquad{}+\alpha_\Phi\left(\frac{1}{2}g^{ij}g^{k\ell}-g^{ik}g^{j\ell}\right)\p_k\phi\p_\ell\phi+\frac{1}{2}\alpha_\Lambda g^{ij}.
\label{eecee}
\eea
Here we denoted by $\CF$ the spacetime integral of the lowest component of the superpotential Lagrangian density in superspace,
\be
\label{eebsup}
\CF=\int dt\,d^Dx\,\sqrt{g}n\left(\alpha_RR^{(g)}+\alpha_\Phi g^{ij}\p_i\phi\p_j\phi+\alpha_\Lambda\right),
\ee
with $R^{(g)}$ in (\ref{eebsup}) now being the bosonic Ricci scalar of $g_{ij}$.  We have also introduced a slight change of notation in the lapse sector:  From now on, we will use 
\be
\phi=-\log n,
\ee
which is often a more convenient variable than the lapse field $n$ itself.  Also, it is this field $\phi$ which will turn out to be simply proportional to Perelman's dilaton $\hat\phi$.  

The equation of motion obtained from varying $B_{ij}$ in the full action $S$ is
\be
2B_{ij}=\frac{1}{n}\nabla_tg_{ij}-(g_{ik}g_{j\ell}-\tilde\lambda g_{ij}g_{k\ell})\CE^{k\ell}\equiv E_{ij}.
\ee
Here the tensor
\be
g_{ik}g_{j\ell}-\tilde\lambda g_{ij}g_{k\ell}
\ee
is the inverse to the DeWitt metric on the space of metrics
\be
g^{ik}g^{j\ell}-\lambda g^{ij}g^{k\ell},
\label{eedewitt}
\ee
with $\tilde\lambda$ given in terms of $\lambda$ by \cite{lif}
\be
\tilde\lambda=\frac{\lambda}{\lambda D-1}.
\ee

By solving for $B_{ij}$ algebraically, the action now becomes
\be
S\approx \frac{1}{4}\int dt\,d^Dx\sqrt{g}n\,E_{ij}\left(g^{ik}g^{j\ell}-\lambda g^{ij}g^{k\ell}\right)E_{k\ell}.
\ee
Assuming that the DeWitt metric (\ref{eedewitt}) is weakly positive definite (\ie , positive definite modulo spatial diffeomorphisms), standard arguments of topological quantum field theory will localize the path integral to the minima of the action, which thus requires the validity of the localization equation 
\be
\label{eeloceqg}
E_{ij}=0.
\ee
For our specific superpotential (\ref{eesuperp}), this equation yields the metric flow 
\bea
\label{eeloceqgg}
e^\phi\nabla_tg_{ij}&=&-\alpha_RR_{ij}+\frac{\alpha_R}{2}\left[1+(2-D)\tilde\lambda\right]g_{ij}R-\alpha_R\nabla_i\p_j\phi\nonumber\\&&\qquad{}+\alpha_R\left[1+(1-D)\tilde\lambda\right]g_{ij}\Delta\phi+(\alpha_R-\alpha_\Phi)\p_i\phi\p_j\phi\\
&&{}+\left\{\frac{\alpha_\Phi}{2}\left[1+(2-D)\tilde\lambda\right]-\alpha_R\left[1+(1-D)\tilde\lambda\right]\right\}g_{ij}(\p\phi)^2+\frac{\alpha_\Lambda}{2}(1-\tilde\lambda D)g_{ij}.\nonumber
\eea
Our next challenge is to identify for which values of the couplings, if any, these localization equations are related to Perelman's Ricci flow.

\subsection{Finding Perelman's equations: The $\diff_\SF(\CM)$ equivariant case}
\label{secff}

Our first step is to propose a change of variables, with the metric rescaled via
\be
\hat g_{ij}=e^\phi g_{ij}.
\label{eeframe}
\ee
This change of frames of the spatial metric is designed to eliminate the extra factor of $e^\phi$ between the two sides of (\ref{eesch}) or (\ref{eeloceqgg}), so that the leading terms on both sides match those of the flow equation (\ref{eepone}).  Given the importance of such reframing transformations throughout our paper, we have collected the relevant change-of-frame formulas for various geometric objects in Appendix~\ref{ssapp} for convenience and completeness.  

Next we need to determine how Perelman's dilaton $\hat\phi$ should be related to our $\phi$.  Rewriting our superpotential $S_\CW$ in the mixed variables $\hat g_{ij}$ and $\phi$ and comparing to Perelman's $\hat\CF$-functional then suggests that we need to set
\be
\hat\phi=\frac{D}{2}\phi.
\label{eerefrphi}
\ee
Finally we also identify $\hat n^i=n^i$.

It is natural to introduce the couplings $\hat\alpha_R$ and $\hat\alpha_\Phi$ in Perelman's functional,
\be
\hat\CF=2\int d^D x\sqrt{\hat g}e^{-\hat\phi}\left\{\hat\alpha_R\,\hat R+\hat\alpha_\Phi\,\hat g^{ij}\p_i\hat\phi\p_j\hat\phi\right\},
\ee
noting that the original $\hat\CF$ functional (\ref{eeperelf}) corresponds to the specific choice of 
\be
\label{eetilal}
\hat\alpha_R=\hat\alpha_\Phi=2.
\ee
Using the reframing formulas from Appendix~\ref{ssapp} shows that in our original variables, this choice translates into
\be
\label{eeal}
\alpha_R=2,\qquad\alpha_\Phi=\frac{2-D}{2}.
\ee
These are the predictions for the regime of our topological gravity where we may expect contact with Perelman's flow equations. 

With this proposed relation (\ref{eeframe}) and (\ref{eerefrphi}) between Perelman's variables and ours, it is now illuminating to rewrite the covariant form of his volume-fixing condition (\ref{eefvol}) in our variables.  It is pleasing to see that this condition becomes simply
\be
\nabla_t\sqrt{g}=0,
\ee
the condition of the spatial volume element being covariantly constant in time.  This observation is very suggestive:  It is indeed well-known in nonrelativistic quantum gravity of the Lifshitz type \cite{mqc,lif,gen,grx} how to realize such a spatial ``unimodularity'' condition dynamically!  This condition is the result of the equations of motion when we take the limit of
\be
|\lambda|\to\infty.
\ee
This regime of nonrelativistic quantum gravity is particularly interesting for a number of reasons \cite{mqc,lif,gen,grx}, and it has also been studied in the context of physical cosmology \cite{shinjic}.  Leaving such physical motivations aside, it is intriguing to see that this same regime of ``gravity at the farpoint'' $\lambda=\pm\infty$ in the kinetic coupling $\lambda$ makes an independent appearance in the topological quantum gravity of Perelman's Ricci flow!

We can now verify how our localization equation (\ref{eeloceqgg}) is precisely related to Perelman's equations.  Taking the values of the couplings in (\ref{eeal}), setting $\lambda=\pm\infty$, and rewriting our localization equations in Perelman's variables using the reframing formulas from Appendix~\ref{ssapp}, our (\ref{eeloceqgg}) becomes
\be
\label{eecovpf}
\hat\nabla_t\hat g_{ij}-\frac{2}{D}\hat g_{ij}\hat\nabla_t\hat\phi=-2\hat R_{ij}-2\hat\nabla_i\p_j\hat\phi+\frac{2}{D}\hat g_{ij}\hat R+\frac{2}{D}\hat g_{ij}\hat\Delta\hat\phi,
\ee
where $\hat\nabla_t\hat\phi\equiv\p_t\hat\phi-\hat n^i\p_i\hat\phi$.  This is just the sum of Perelman's first equation (\ref{eepone}) and $-(D/2)\hat g_{ij}$ times Perelman's second equation (\ref{eeptwo})!  Note that we did not have to take the cosmological constant $\alpha_\Lambda$ to zero:  This is one of the benefits of being at the ``farpoint'' $|\lambda|=\infty$ in the kinetic coupling $\lambda$.  Moreover, by taking the trace of (\ref{eecovpf}), we obtain 
\be
\hat g^{ij}\hat\nabla_t\hat g_{ij}-2\hat\nabla_t\hat\phi=0,
\ee
which can be usefully rewritten as
\be
\hat\nabla_t\left(e^{-\hat\phi}\sqrt{\hat g}\right)=0.
\ee
This confirms that in accord with our anticipation, our theory in the $|\lambda|=\infty$ limit indeed dynamically imposes Perelman's fixed-volume condition (\ref{eefvol}), in its covariant form, and in a form in which the rather awkward and unnecessary reference to an arbitrary constant measure $dm(x^i)$ on $\Sigma$ in (\ref{eefvol}) is nicely absent.

This is as close as we can get to Perelman's original equations in the covariant theory studied in this section.  The two equations cannot be separated in a covariant way:  Such a split would be inconsistent with the fact that the localization equations in our theory with the $\CG$ gauge symmetry realized equivariantly must be gauge invariant under time-dependent time reparametrization, a symmetry not respected by the individual equations (\ref{eepone}) and (\ref{eeptwo}) but respected by their appropriate sum.  Thus, our system (\ref{eecovpf}) represents a covariantized version of Perelman's equations, made consistent with time reparametrizations.

\subsection{Gauge fixing of spatial diffeomorphisms and DeTurck's trick}
\label{secdtt}

The theory is still gauge invariant under the symmetry of spacetime-dependent spatial diffeomorphisms.  We can perform a gauge fixing of this symmetry to make contact with Perelman's equations before and after the DeTurck trick.  The simplest natural gauge choice, which we referred to in \cite{grf} as \textit{Perelman gauge}, is to set $n^i=0$.  In this case, we simply obtain Perelman's original flow equations (\ref{eepone}) and (\ref{eeptwo}).

Another natural gauge fixing, which we referred to in \cite{grf} as \textit{Hamilton gauge}, is given in Perelman's variables by
\be
\hat n_i=\p_i\hat\phi,
\ee
where we define $\hat n_i\equiv\hat g_{ij}n^j$.  In our original variables, this condition can be rewritten as
\be
e^\phi n_i=\frac{D}{2}\p_j\phi,\quad\mathrm{or}\quad n_i=-\frac{D}{2}\p_in.
\ee
With this choice of $n_i$, our localization equations in Perelman's variables become
\be
\dot{\hat g}_{ij}-\frac{2}{D}\hat g_{ij}\dot{\hat\phi}=-2\hat R_{ij}+\frac{2}{D}\hat g_{ij}\left(\hat R+\hat\Delta\hat\phi-\hat g^{k\ell}\p_k\hat\phi\p_\ell\hat\phi\right),
\ee
which nicely reproduces the sum of Perelman's equations after the DeTurck trick, (\ref{eepdtone}) and (\ref{eepdttwo}).

Having shown how to use spatial diffeomorphisms to perform DeTurck's trick as their gauge fixing, one might still be concerned that our theory contains $D(D+1)/2$ localization equations (\ref{eeloceqgg}) for $[D(D+1)/2]+1$ field variables $g_{ij}$ and $\phi$, given the fact that our lapse sector is nonprojectable and $\phi$ therefore spacetime-dependent.  Where is the missing equation, which would provide the localization condition in the $\phi$ sector?

In Type C theory, on which we have focused in this section so far, the fact that we are missing one localization equation is consistent:  By imposing the chirality condition on the lapse superfield $E$, we effectively set the antighost and the auxiliary field associated with $\phi$ to zero.  Thus, our Type C theory is not yet fully gauge fixed: For example, the ghost field in the lapse sector does not have a non-degenerate kinetic term.  Choosing a BRST-trivial antighost-auxiliary multiplet and adding another gauge-fixing term to the action would be required in order to complete the construction of the theory.  

In Type B theory \cite{grf} on the other hand, these remaining gauge-fixing ingredients are already present.  Since one does not impose any chirality condition on the lapse superfields, the component content is then ``balanced'' (in the sense reviewed in \cite{grf}):  $\nu$ and $\bar\nu$ are the ghost and the antighost, and $w$ is a bosonic auxiliary.  Note that the secondary gauge symmetry $\CG$ that we require throughout this section only contains \textit{spatially-independent} time reparametrizations, a symmetry clearly not large enough to mimic what we did in the shift sector and set these three component fields to zero by some Wess-Zumino-type gauge.  If we choose the same action (\ref{eecovsk}) and (\ref{eesuperp}) in the Type B theory, the $w$ auxiliary field will yield the missing scalar flow equation.   It is intriguing that it does so in a time-reparametrization covariant way.

This last observation also clearly indicates that the Type B theory, with the gauge symmetry $\CG$ treated equivariantly, will \textit{not} be the correct setting if our goal is to obtain Perelman's Ricci flow as the exact localization equations.  In order to achieve that goal, we will have to revisit the original gauge symmetries and their gauge fixing, following the arguments developed in \cite{gst}.  This will be the main focus of Section~\ref{secpf} below.  However, since the Type B theory with the gauge symmetry $\CG$ may still be of independent interest, we present the structure of its localization equations now, before we proceed with our main task in Section~\ref{secpf}.

\subsection{Type B theory}

In Type B theory, the bosonic part of the action is
\be
S_K\approx-\int dt\,d^Dx\,\sqrt{g}n\,(g^{ik}g^{j\ell}-\lambda g^{ij}g^{k\ell})\,\CB_{ij}\CB_{k\ell}
+\int dt\,d^Dx\,\sqrt{g}\,(g^{ik}g^{j\ell}-\lambda g^{ij}g^{k\ell})\,\CB_{ij}\nabla_tg_{k\ell},
\ee
and
\be
S_\CW\approx\int dt\,d^Dx\sqrt{g}n\,\CB_{ij}\CE^{ij}+\int dt\,d^Dx\,w\pi,
\ee
where $\CE^{ij}$ is again given by (\ref{eecee}), and 
\bea
\pi&=&\alpha_R\sqrt{g}\left(g^{ik}g^{j\ell}-g^{ij}g^{k\ell}\right)\nabla_i\left(\vphantom{g^{ij}}\nabla_jn\nabla_tg_{k\ell}-n\nabla_j\nabla_tg_{k\ell}\right)\nonumber\\
&&\qquad{}-2\alpha_\Phi \p_i\left[n\nabla_t\left(\sqrt{g}g^{ij}\p_j\phi\right)\right].
\eea
In order to get the component action into such a nice diagonalized form in the auxiliaries, we had to perform a simple redefinition of the auxiliary fields $B_{ij}$ and $w$,
\be
\CB_{ij}=B_{ij}-w\nabla_t g_{ij}.
\ee
This field transformation from $B_{ij}, w$ to $\CB_{ij},w$ has the unit Jacobian and therefore it is an allowed change of variables in the path integral, including its component field measure.

The equations of motion that are obtained from varying $\CB_{ij}$ and $w$ in the full action $S$ are:
\bea
2\CB_{ij}&=&\frac{1}{n}\nabla_tg_{ij}-(g_{ik}g_{j\ell}-\tilde\lambda g_{ij}g_{k\ell})\CE^{k\ell}\equiv E_{ij},\\
\pi&=&0.
\eea
Here $E_{ij}$ is again the same as in Type C theory above.  The path integral over $w$ yields a delta function and imposes the constraint $\pi=0$, which represents the ``missing'' scalar equation.  Unfortunately, we have not been able to decode the geometric meaning of this $\pi=0$ constraint in terms relevant to the mathematical theory of the Ricci flow equations.  

\section{Perelman's equations and gauge fixing of time reparametrizations}
\label{secpf}

We are now ready to combine together the pieces of the puzzle that we learned in \cite{grf,gst} and in this paper so far, and finally to write down the precise topological quantum gravity theory whose localization equations are equivalent to Perelman's Ricci flow, after the appropriate change of frames.

\subsection{The theory}

The key is to choose the suitable underlying gauge symmetry, and to gauge fix it such that the residual symmetries match the symmetries of Perelman's equations.  The crucial lesson was learned in \cite{gst}, where we showed how the superspace construction of \cite{grf} can be interpreted as the one-step BRST gauge fixing of a theory of the ADM metric variables, with a non-redundant gauge symmetry.  In \cite{gst}, this symmetry was a combination of topological deformations of the spatial metric $g_{ij}$ and the ultralocal limit of all spacetime diffeomorphisms acting on the lapse and shift $n$ and $n^i$.  It will be useful to change the gauge symmetry to a closely related, also non-redundant symmetry $\SG$, generated by
\bea
\delta n&=&f(t,x^i),\\
\delta n^i&=&\dot\xi^i+\xi^k\p_kn^i-n^k\p_k\xi^i,\\
\delta g_{ij}&=&f_{ij}(t,x^k).
\eea
The interpretation of this symmetry structure is as follows:  $\xi^i$ acts via standard spacetime-dependent spatial diffeomorphisms on $n^i$ as indicated, and on $n$ and $g_{ij}$ in the standard way as well.  In addition, we have arbitrary topological deformations $f_{ij}$ of the spatial metric $g_{ij}$, as well as arbitrary topological deformations $f$ of the lapse $n$; much like in \cite{gst}, we have absorbed the action by $\xi^i$ on $g_{ij}$ and $n$ into a shift in the definition of $f$ and $f_{ij}$, certainly a change of variables whose Jacobian is equal to one.  This explains the simplicity of the gauge transformations, and makes it clear that the gauge symmetries are non-redundant.  Note that this symmetry structure leads to zero local propagating degrees of freedom, and thus a topological theory.  

We choose to realize only the spacetime-dependent spatial diffeomoprhisms equivariantly.  The symmetries generated by $f$ and $f_{ij}$ need to be gauge fixed.  Thus, the $\CN=2$ BRST superfields that we will use are as follows.  In the metric sector, we use the same unconstrained metric superfield $G_{ij}$ that we have used so far,
\be
G_{ij}=g_{ij}+\theta\psi_{ij}+\bar\theta\chi_{ij}+\theta\bar\theta\,B_{ij}.
\ee
The superfields $N^i$, $S^i$ and $\bar S^i$ in the shift sector will be identical to those used in \cite{grf,gst} and in this paper; their component fields are $n^i$, $\sigma^i$, $\bar\sigma^i$ and $X^i$, interpreted as the shift vector, its ghost, its antighost, and its auxiliary associated with the gauge symmetries generated by $\xi^i$.  We will again adopt the Wess-Zumino gauge, setting $\sigma^i$, $\bar\sigma^i$ and $X^i$ to zero.
Finally, in the lapse sector we will use an unconstrained lapse superfield $N$, or better yet, the superfield $\Phi\equiv -\log N$, whose lowest component is our $\phi$:%
\footnote{Note that this is the only place where our conventions in this paper differ from those of \cite{gst}, where $\psi$ and $\chi$ were used to denote the component fermions of the $E$ superfield, $E=\log\Phi$.}
\be
\Phi=\phi+\theta\psi+\bar\theta\chi+\theta\bar\theta\SB.
\ee

Given these component fields, the action of the gauge-fixed theory will be of the form
\be
S=\int dt\,d^Dx\{Q,\Psi\},
\label{ssactpsi}
\ee
with an appropriately chosen gauge-fixing fermion $\Psi$.  Writing this action as a superspace integral
\be
S=\int dt\,d^Dx\,d^2\theta\,\CL
\ee
with some superspace Lagrangian $\CL$ will make sure that the action is of the form (\ref{ssactpsi}) for some gauge-fixing fermion $\Psi$, and it will also ensure the extension to the $\CN=2$ BRST supersymmetry.  Now, everything rests on the choice of $\Psi$.  

It is the entire idea of gauge fixing to make sure that the choice of the gauge-fixing fermion $\Psi$ fixes the part of the gauge group that we wish to gauge-fix.  In our case, we decided to keep the spatial diffeomorphisms generated by $\xi^i$ unfixed, so that we can make contact with the symmetries of the Ricci flow.  Thus, $\Psi$ (or, equivalently, $\CL$) must be chosen such that it fixes the topological symmetries generated by $f$ and $f_{ij}$, while still being invariant under $\xi^i$.  This is where our construction will be different from that in Section~\ref{seccpf} above:  The gauge-fixing fermion of the $\diff_\SF(\CM)$ covariant theory studied in Section~\ref{seccpf} was chosen such that the time-dependent time diffeomorphisms were unfixed; here we choose $\Psi$ such that even these residual time reparametrizations are gauge-fixed.

The appropriate modification will come from the kinetic sector.  First, instead of the minimal kinetic term (\ref{eecovsk}) with time-reparametrization gauge invariance, we now begin with 
\be
S^{(0)}_K=\int d^2\theta\,dt\,d^Dx\,\sqrt{G}\,(G^{ik}G^{j\ell}-\lambda G^{ij}G^{k\ell})\,\bar\CD G_{ij}\,\CD G_{k\ell}.
\ee
The superderivatives $\CD$ and $\bar\CD$ are covariant with respect to spatial diffeomorphisms, but not with respect to any time reparametrizations.  Their definition is the same as in \cite{grf},
\bea
\CD G_{ij}&\equiv& \RD G_{ij}-S^k\p_kG_{ij}-G_{kj}\p_iS^k-G_{ik}\p_jS^k,\\
\bar\CD G_{ij}&\equiv& \barRD G_{ij}-\bar S^k\p_kG_{ij}-G_{kj}\p_i\bar S^k-G_{ik}\p_j\bar S^k.
\eea
By design, this kinetic term fixes not just the topological gauge symmetry of $g_{ij}$ but also the gauge symmetries acting on $n$, and leaves no local time reparametrizations unfixed.  To this kinetic term, we are free to add other terms of the same classical scaling dimension that are of the minimal form in derivatives and respect the same symmetries:  We can add terms of the form $\CD\Phi\,\bar\CD\Phi$, $G^{ij}\CD G_{ij}\,\bar\CD\Phi$ and $G^{ij}\bar\CD G_{ij}\,\CD\Phi$, with independent couplings.  For our purposes, the first one -- the kinetic term for $\Phi$ -- will be sufficient, and the couplings of the off-diagonal terms mixing $\Phi$ with the metric will be set to zero.  Thus, our full kinetic term will be
\be
S_K=\int d^2\theta\,dt\,d^Dx\,\sqrt{G}\left\{\,(G^{ik}G^{j\ell}-\lambda G^{ij}G^{k\ell})\,\bar\CD G_{ij}\,\CD G_{k\ell}+\lambda_\Phi\,\bar\CD\Phi\,\CD\Phi\right\}.
\label{eefullk}
\ee
As we will see below, the value of the coupling constant $\lambda_\Phi$ for which the match to Perelman's equations will be accomplished is $\lambda_\Phi=D$.  The superpotential (\ref{eesuperp}) stays unchanged,
\be
S_\CW=\int d^2\theta\,dt\,d^Dx\,\sqrt{G}e^{-\Phi}\left\{\alpha_RR^{(G)}+\alpha_\Phi G^{ij}\p_i\Phi\p_j\Phi+\alpha_\Lambda\right\},
\label{eefullw}
\ee
except that now $\Phi$ is an unconstrained superfield, with the lapse also unconstrained and given by $N=\exp(-\Phi)$.

In components, the bosonic part of the action is
\bea
S&\approx& -\int dt\,d^Dx\,\sqrt{g}\left\{B_{ij}\left(g^{ik}g^{j\ell}-\lambda g^{ij}g^{k\ell}\right)B_{k\ell}+\lambda_\Phi\SB^2\right\}\nonumber\\
&&{}+\int dt\,d^Dx\,\sqrt{g}\,B_{ij}\left\{\left(g^{ik}g^{j\ell}-\lambda g^{ij}g^{k\ell}\right)\nabla_t g_{k\ell}-\SE^{ij}\right\}\\
&&\qquad{}-\int dt\,d^Dx\,\sqrt{g}\,\SB\left(\lambda_\Phi \nabla_t\phi-\SE\right),\nonumber
\eea
where $\nabla_t$ continues to denote the time derivative covariantized with respect to spatial diffeomorphisms,
\bea
\nabla_t g_{ij}&=&\dot g_{ij}-\nabla_in_j-\nabla_jn_i,\\
\nabla_t\phi&=&\dot\phi-n^k\p_k\phi,
\eea
and where
\bea
\SE^{ij}\equiv\frac{1}{\sqrt{g}}\frac{\delta\CF}{\delta g_{ij}}&=&e^{-\phi}\left\{\alpha_R\left(-R^{ij}+\frac{1}{2}Rg^{ij}\right)+\left(\frac{1}{2}\alpha_\Phi-\alpha_R\right)g^{ij}(\p\phi)^2+\alpha_Rg^{ij}\Delta\phi\right.\nonumber\\
&&\left.\quad{}+(\alpha_R-\alpha_\Phi)g^{im}g^{js}\p_m\phi\p_s\phi-\alpha_Rg^{im}g^{js}\nabla_m\p_s\phi+\frac{\alpha_\Lambda}{2}g^{ij}\vphantom{\frac{1}{2}}\right\},\\
\SE\equiv\frac{1}{\sqrt{g}}\frac{\delta\CF}{\delta\phi}&=&e^{-\phi}\left\{-\alpha_RR+\alpha_\Phi\left[g^{ij}\,\p_i\phi\,\p_j\phi-2\Delta\phi\right]-\alpha_\Lambda\right\}, 
\eea
with $\CF$ again given by
\be
\CF=\int dt\,d^Dx\,\sqrt{g}e^{-\phi}\left\{\alpha_RR^{(g)}+\alpha_\Phi g^{ij}\p_i\phi\p_j\phi+\alpha_\Lambda\right\}.
\ee
In the process of deriving $\delta\CF/\delta g_{ij}$, it is important to recall the correct formula for the variation of the Einstein-Hilbert scalar curvature term,
\be
\delta(\sqrt{g} R)=-\sqrt{g}\left(R^{ij}-\frac{1}{2}Rg^{ij}\right)\delta g_{ij}+\sqrt{g}\left(g^{ik}g^{j\ell}-g^{ij}g^{k\ell}\right)\nabla_i\nabla_j\delta g_{k\ell}.
\ee
It then follows from the form of our component action that the localization equations are
\bea
e^\phi\nabla_t g_{ij}&=&e^\phi\left(g_{ik}g_{j\ell}-\tilde\lambda g_{ij}g_{k\ell}\right)\SE^{k\ell}\equiv -\alpha_RR_{ij}+\frac{\alpha_R}{2}\left[1-\tilde\lambda(D-2)\right]g_{ij}R\nonumber\\
&&{}+(\alpha_R-\alpha_\Phi)\p_i\phi\p_j\phi+\left[\left(\frac{\alpha_\Phi}{2}-\alpha_R\right)(1-\tilde\lambda D)+(\alpha_\Phi-\alpha_R)\tilde\lambda\right]g_{ij}(\p\phi)^2
\nonumber\\
&&\qquad\qquad{}+\alpha_R\left[1-\tilde\lambda(D-1)\right]g_{ij}\Delta\phi-\alpha_R\nabla_i\p_j\phi+\frac{\alpha_\Lambda}{2}(1-\tilde\lambda D)g_{ij},
\label{eelocal1}\\
\lambda_\Phi e^\phi\nabla_t\phi&=&-\alpha_RR+\alpha_\Phi\left[g^{ij}\,\p_i\phi\,\p_j\phi-2\Delta\phi\right]-\alpha_\Lambda.
\label{eelocal2}
\eea
We see that in this theory, the previously ``missing'' localization equation for $\phi$ has been supplied by our choice of the gauge-fixing fermion leading to the action (\ref{eefullk}) and (\ref{eefullw}).  

\subsection{Reframing to Perelman's variables}

The main conclusion from our investigations of the covariant theory in Section~\ref{seccpf} was that our variables $g_{ij}$ and $\phi$ are related to Perelman's variables $\hat g_{ij}$ and $\hat\phi$ by
\bea
\label{eerefmet}
\hat g_{ij}&=&e^\phi g_{ij},\\
\hat\phi&=&\frac{D}{2}\phi.
\eea
In addition, the shift vector transforms trivially, $\hat n^i=n^i$.  We will accept these relations here as well.  Its strongest motivation comes from our desire to see Perelman's volume condition take the simple form of a covariant constancy of $\sqrt{g}$ in our variables, so that we can dynamically impose it by taking the $|\lambda|\to\infty$ limit.

\begin{figure}[t!]
  \centering
    \includegraphics[width=0.75\textwidth]{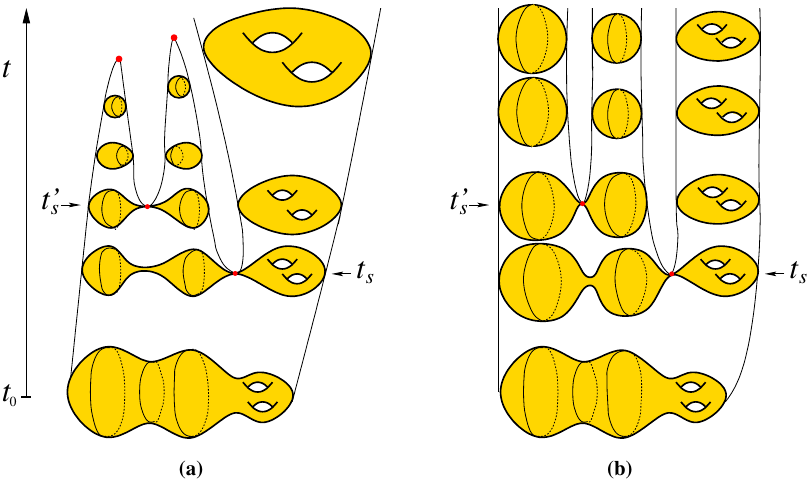}
    \caption{A typical qualitative example of a solution of Perelman's Ricci flow equations in $3+1$ dimensions, which includes both topology-changing transitions at time instants $t_s$ and $t'_s$ and Perelman's extinction of positively curved spatial regions. \textbf{(a):} The evolution in the Perelman frame, and \textbf{(b):} in our frame.}
    \label{ff}
\end{figure}

As an aside remark, we note that the reframing transformation (\ref{eerefmet}) from Perelman's metric to ours will have an interesting effect on the solutions of Perelman's Ricci flow equations.  Consider the case of $3+1$ dimensions, for which the most detailed information is available in the mathematical literature.  As we briefly reviewed in the Introduction to \cite{grf}, under the influence of Perelman' Ricci flow, spatial geometries with positive sectional curvatures round themselves out with time and shrink to an extinguishing singularity in finite time.  In contrast, hyperbolic spatial geometries with negative sectional curvatures expand forever.  Besides the extinguishing singularities of positively-curved regions, the geometries can also go through topology-changing ``neckpinch'' singularities.  We have illustrated such a generic evolution of an initial spatial geometry in Figure~\ref{ff}(a).  All these features are found when the spatial metric is in Perelman's frame, $\hat g_{ij}$.  

The reframing to our frame $g_{ij}$ has an interesting effect on the qualitative behavior of the solutions.  Viewed in our frame, the positively-curved regions round themselves up, but approach a constant radius limit asymptotically, as $t\to\infty$.  Similarly, the metric of the hyperbolic regions also approaches a stationary limit as $t\to\infty$.  In contrast, the topology-changing ``neckpinch'' singularities still happen at finite time.  We illustrate this qualitative behavior in our frame in Figure~\ref{ff}(b).

This contrast between the spacetime geometry viewed from different frames does not mean that the spacetime configuration is somehow different between the two frames:  It just shows, in the context of the Ricci flow, that viewing the same geometric solution in different frames may reveal new features, often difficult to see if one insists on one preferred frame.  This phenomenon is well-understood in string theory, where the same spacetime geometry can be probed by different probes (such as strings, or branes of various dimensions), revealing complementary information about the same solution of the theory.  It is pleasing to see a similar behavior in the topological quantum gravity of the Ricci flow.

\subsection{Gravity at farpoint: Taking the $|\lambda|\to\infty$ limit}

We know that the desired fixed-volume condition in our variables will be dynamically imposed when we take the ``farpoint'' limit of the kinetic coupling $\lambda$, taking
\be
\lambda\to\pm\infty.
\ee
At the level of the localization equations, either of these two limits is permissible: In fact, they both lead to the same localization equations.  Indeed, in terms of the dual coupling $\tilde\lambda$, both cases correspond to the same value,
\be
\tilde\lambda=\frac{1}{D}.
\ee
While the localization equations are formally identical in both limits $\lambda\to\pm\infty$, the two cases differ from the perspective of the path integral.  To see that, let us integrate out the auxiliary fields $B_{ij}$ and $\SB$ in the path integral.  The reduced action is now a sum of squares of the localization equations, 
\be
S\approx\int dt\,d^Dx\sqrt{g}\left\{\frac{1}{4}\left(g^{ik}g^{j\ell}-\lambda g^{ij}g^{k\ell}\right)(\nabla_t g_{ij}-\ldots\,)(\nabla_t g_{k\ell}-\ldots\,)+\frac{\lambda_\Phi}{4}(\nabla_t\phi-\ldots\,)^2\right\},
\ee
where the ``$\ldots$'' stand for the right-hand sides of the localization equations determining $\nabla_t g_{ij}$ and $\nabla_t\phi$.  We see that for $\lambda\to-\infty$, and with $\lambda_\Phi>0$, the action is manifestly $S\geq 0$, with the equality saturated exactly for those bosonic configurations that satisfy the flow equations.   This is then the preferred value to take, in order to make the path integral well-defined.  Taking the other limit, $\lambda\to\infty$, would put us in a situation similar to the one encountered in general relativity, where the Euclidean action is not bounded from below and the path integral requires a subtle analytic continuation.  

When the $|\lambda|\to\infty$ limit is taken, our localization equations (\ref{eelocal1}) and (\ref{eelocal2}) reduce to the more manageable system
\bea
e^\phi\nabla_t g_{ij}&=&
-\alpha_RR_{ij}+\frac{\alpha_R}{2}g_{ij}R+\frac{\alpha_R}{D}g_{ij}\Delta\phi-\alpha_R\nabla_i\p_j\phi\nonumber\\
&&\qquad{}+(\alpha_R-\alpha_\Phi)\p_i\phi\p_j\phi+\frac{1}{D}(\alpha_\Phi-\alpha_R)g_{ij}(\p\phi)^2,
\label{eellm}\\
\lambda_\Phi e^\phi\nabla_t\phi&=&-\alpha_RR+\alpha_\Phi\left[g^{ij}\,\p_i\phi\,\p_j\phi-2\Delta\phi\right]-\alpha_\Lambda.
\label{eelln}
\eea
Having determined our full list of localization equations in the ``farpoint'' limit of $|\lambda|\to\infty$, and having committed to the change-of-frame relation between our variables and Perelman's, we can now determine whether Perelman's Ricci flow corresponds to a particular choice of our couplings $\alpha_R$, $\alpha_\Phi$ and $\alpha_\Lambda$ after reframing.

\subsection{Localization and Perelman's Ricci flow}

We consider the covariantized form of Perelman's flow equations,
\bea
\label{eeponec}
\hat\nabla_t\hat g_{ij}&=&-2\hat R_{ij}-2\hat\nabla_i\p_j\hat\phi,\\
\label{eeptwoc}
\hat\nabla_t\hat\phi&=&-\hat R-\hat\Delta\hat\phi.
\eea
In Perelman gauge $\hat n^i=0$, these reduce back to Perelman's original system (\ref{eepone}) and (\ref{eeptwo}).  We begin with the scalar flow equation (\ref{eeptwoc}) for $\hat\phi$.  Does it match our scalar localization equation (\ref{eelln}) after reframing?  This is a nontrivial check:  Notice that both the $\Delta\phi$ term and the $(\p\phi)^2$ term on the right-hand side of our (\ref{eelln}) are controlled by the \textit{same} coupling $\alpha_\Phi$.  If the reframing of the right-hand side of Perelman's scalar flow equation (\ref{eeptwoc}) yields the two terms $\Delta\phi$ and $(\p\phi)^2$ with a relative coefficient different than $-2$, our program would fail.  Happily, applying our reframing formulas from Appendix~\ref{ssapp} gives
\be
\hat R+\hat\Delta\hat\phi
=e^{-\phi}\left\{R+\frac{D-2}{4}\left[(\p\phi)^2-2\Delta\phi\right]\right\}.
\label{eeq}
\ee
Comparing this to our scalar flow (\ref{eelln}), we obtain the ratio of $\alpha_R$ and $\alpha_\Phi$ in our frame's Lagrangian,
\be
\alpha_\Phi=\frac{2-D}{4}\alpha_R.
\label{eephir}
\ee
This is indeed the same ratio of the couplings which was predicted in (\ref{eeal})!
In fact, we can set $\alpha_R=2$ by convention, and use (\ref{eephir}) to determine that
$\alpha_\Phi=(2-D)/2$, as given in (\ref{eeal}).  We must also set the cosmological constant $\alpha_\Lambda=0$. Finally, we see that the scalar flows will match exactly if we set $\lambda_\Phi=D$.

With these values of the couplings, it remains to see whether our metric flow equation matches Perelman's after reframing.  There is no more freedom of choice of any couplings left, so this highly nontrivial check in the metric sector must work identically, for our goal to succeed.  The right-hand side of our metric flow equation (\ref{eellm}) now simplifies to
\be
e^\phi\nabla_t g_{ij}=-2R_{ij}+\frac{2}{D}g_{ij}R-\frac{2+D}{2D}g_{ij}(\p\phi)^2+\frac{2+D}{2}\p_i\phi\p_j\phi+\frac{2}{D}g_{ij}\Delta\phi-2\nabla_i\p_j\phi.
\label{eeourm}
\ee
Note that this expression is correctly traceless, as it must, since it corresponds to the $\lambda=\pm\infty$ limit which imposes dynamically the unimodularity condition on $g_{ij}$.  This is a good check of self-consistency of our framework.  

The question is whether (\ref{eeourm}) is a reframing of the right-hand side of the Perelman flow equation for the metric.  Let us begin with the right-hand side of Perelman's flow equation for the metric:
\be
2\hat R_{ij}+2\hat\nabla_i\p_j\hat\phi.
\label{eepermet}
\ee
In order to write this expression in our frame, we need to invoke the reframing equation (\ref{eeuncontr}) from the Appendix.  Using this relation, we find that the reframing of (\ref{eepermet}) is
\be
2R_{ij}+2\nabla_i\p_j\phi-g_{ij}\Delta\phi-\frac{D+2}{2}\p_i\phi\p_j\phi+g_{ij}(\p\phi)^2.
\ee
This does not look at all like the right-hand side of (\ref{eeourm}).  However, it should not yet look like it, because (\ref{eepermet}) is the right-hand side of Perelman's equation for \textit{his} metric $\hat g_{ij}$, while (\ref{eeourm}) is the right-hand side of the flow equation for \textit{our} metric $g_{ij}$.  Since they are related by $\hat g_{ij}=e^\phi g_{ij}$, we have
\be
\hat\nabla_t\hat g_{ij}=e^\phi\left(\nabla_t g_{ij}+g_{ij}\nabla_t\phi\right).
\ee
Hence, to compare (\ref{eepermet}) to (\ref{eeourm}), we must subtract from it $e^\phi g_{ij}\dot\phi$, and substitute for $e^\phi\dot\phi$ using our scalar flow equation whose right-hand side is in (\ref{eeq}).  This gives
\bea
2R_{ij}&+&2\nabla_i\p_j\phi-g_{ij}\Delta\phi-\frac{D+2}{2}\p_i\phi\p_j\phi+g_{ij}(\p\phi)^2\nonumber\\
&-&\frac{2}{D}g_{ij}\left\{R+\frac{D-2}{4}\left[(\p\phi)^2-2\Delta\phi\right]\right\}.
\eea
Simple algebra then shows that this expression is
\be
2R_{ij}-\frac{2}{D}g_{ij}R+2\nabla_i\p_j\phi-\frac{2}{D}g_{ij}\Delta\phi-\frac{D+2}{2}\p_i\phi\p_j\phi+\frac{D+2}{2D}g_{ij}(\p\phi)^2.
\ee
Happily, this indeed coincides with our $-\SE_{ij}$ in (\ref{eeourm}).  This concludes the proof that the localization equations of our topological quantum gravity, for the values of the couplings
\be
\alpha_R=2,\quad \alpha_\Phi=\frac{2-D}{2},\quad \alpha_\Lambda=0,\quad \lambda=\pm\infty,\quad \lambda_\Phi=D,
\ee
given by
\bea
e^\phi\nabla_t g_{ij}&=&-2R_{ij}+\frac{2}{D}Rg_{ij}-2\nabla_i\p_j\phi+\frac{2}{D}\Delta\phi g_{ij}+\frac{D+2}{2}\p_i\phi\p_j\phi-\frac{D+2}{2D}(\p\phi)^2g_{ij},\ \ 
\label{eeponeour}\\
e^\phi\nabla_t\phi&=&-\frac{2}{D}R-\frac{D-2}{2D}\left[(\p\phi)^2-2\Delta\phi\right],
\label{eeptwoour}
\eea
are exactly equivalent to Perelman's Ricci flow equations (\ref{eeponec}) and (\ref{eeptwoc}), after the appropriate change of frames.  This is the central result of the present paper.

\section{Shrinking and expanding solitons: Perelman's $\CW$ entropy functional}
\label{sectime}

While Perelman's $\hat\CF$-functional is at the core of the modern theory of the Ricci flow, it turns out to be best suited for the study of static solitons.  The important cases of shrinking and expanding solitons are associated with refined versions of the $\hat\CF$ functional, known as the $\CW$ and $\CW_+$ entropy functionals.  These satisfy important monotonicity properties along the appropriate Ricci flow.

\subsection{Shrinking solitons and the $\CW$ entropy functional}

In his analysis of shrinking solitons \cite{perel1}, Perelman introduced the entropy $\CW$-functional,
\be
\CW=\int d^D x\sqrt{\hat g}\frac{1}{(4\pi\tau)^{D/2}}e^{-\hat\phi}\left\{\tau\left(\hat R+\hat g^{ij}\p_i\hat\phi\p_j\hat\phi\right)+\hat\phi -D\right\},
\ee
which depends -- besides the fields $\hat g_{ij}$ and $\hat\phi$ -- on a projectable field $\tau(t)$.  In the original context of \cite{perel1}, this field was playing the role of a spatial scale, related to Perelman's original inspiration from the renormalization group behavior of non-linear sigma models in string theory.  We shall comment on the interpretation of this field $\tau$ in our topological gravity below.

Using the $\CW$-functional instead of the $\CF$-functional in the variational principle, and imposing a modified fixed-volume condition which requires that
\be
\label{eewvol}
\frac{1}{(4\pi\tau)^{D/2}}e^{-\hat\phi}\sqrt{\hat g}\,d^D x 
\ee
be held fixed with time, one obtains the modified gradient flow equations (see Ch.~6.1 of \cite{rf1} for detailed derivation),
\bea
\label{eepfw}
\frac{\p\hat g_{ij}}{\p t}&=&-2\hat R_{ij}-2\hat\nabla_i\p_j\hat\phi,\nonumber\\
\frac{\p\hat\phi}{\p t}&=&-\hat R-\hat\Delta\hat\phi+\frac{D}{2\tau},\\
\frac{\p\tau}{\p t}&=&-1.\nonumber
\eea
The last of these equations is usually solved by setting $\tau=t_0-t$ for some positive constant $t_0$, and then $\tau$ is usually interpreted in the mathematical literature of the Ricci flow as such a linear function of time. 

Just as in the case of the original equations (\ref{eepone}) and (\ref{eeptwo}), these modified equations can be further simplified by DeTurck's trick to
\bea
\label{eepfwdt}
\frac{\p\hat g_{ij}}{\p t}&=&-2\hat R_{ij},\nonumber\\
\frac{\p\hat\phi}{\p t}&=&-\hat R-\hat\Delta\hat\phi+\hat g^{ij}\p_i\hat\phi\p_j\hat\phi+\frac{D}{2\tau},\\
\frac{\p\tau}{\p t}&=&-1,\nonumber
\eea
which again separates Hamilton's original metric flow from the flow of the dilaton.

It is natural to ask whether our construction from Section~\ref{seccpf} can be extended to accommodate the $\CW$-functional and its modified flow equations in our topological quantum gravity.

\subsection{Adding the Goldstone superfield $T$}

In the context of quantum gravity, we must first find an interpretation of Perelman's $\tau$ function as a dynamical field.  In fact, this field can be interpreted in several seemingly distinct ways,%
\footnote{The field $\tau$ can be intepreted as the dilaton for anisotropic conformal transformations \cite{lif,aci} of spacetime (see Ch.~13.2 of \cite{gsw2} for a particularly lucid discussion of the relation between the dilaton and scale invariance); or it can be interpreted as a compensator field similar to those that appear in relativistic supergravity \cite{siegel}.  It can also be interpreted as the Goldstone field associated with spontaneous breaking of time translation symmetries in the background, analogous to a very similar Goldstone field $\pi$ that appears prominently in the effective field theory approach to cosmological inflation \cite{bmca,cremi,cheung,wein1,wein2,wein3}.}
which all stem from the simple idea of spontaneous symmetry breaking.  The symmetry in question is the symmetry of global time translations.  While global time translations may be an isometry of the static Ricci flow solitons most suitable for the $\CF$ functional, shrinking or expanding Ricci solitons (or any other time-dependent background solution) will break this symmetry spontaneously.  On general grounds, it is natural to expect the presence of a gapless Goldstone mode associated with this global symmetry breaking.  A very similar field has played a prominent role in modern cosmology, in the effective field theory of inflation \cite{bmca,cremi,cheung,wein1,wein2,wein3}  (see also \cite{shiu} for further geometric clarifications).  

In order to accommodate Perelman's $\tau$ field consistently with our $\CN=2$ supersymmetry, we first promote it to a superfield.  We choose to introduce a \textit{projectable} superfield $T(t,\theta,\bar\theta)$, otherwise unconstrained, and require its dynamics to be consistent with the condition of a constant shift symmetry, 
\be
T(t,\theta,\bar\theta)\mapsto T(t,\theta,\bar\theta)+ c.
\label{eecshift}
\ee
This shift symmetry is indeed a hallmark of $T$ being a Goldstone field.  While in nonrelativistic systems, shift symmetries allow an intriguing refinement \cite{gra,sur} on symmetric backgrounds, for our purposes it is sufficient to consider this simplest case, which is background independent.

Next, we choose the action of the $T$ sector to be of the minimal form.  The kinetic term is augmented to 
\be
S_K=\int d^2\theta\,dt\,d^Dx\,\sqrt{G}\left\{\,(G^{ik}G^{j\ell}-\lambda G^{ij}G^{k\ell})\,\bar\CD G_{ij}\,\CD G_{k\ell}+\lambda_\Phi\,\bar\CD\Phi\,\CD\Phi\right\}+
\int dt\,d^2\theta\,\barRD T\,\RD T.
\label{eenewsk}
\ee
The lowest-derivative kinetic term for $T$ is indeed consistent with the constant shift symmetry.  For a projectable $T$, we also find that no $T$-dependent terms can be added to the superpotential, which stays the same as before, 
\be
S_\CW=\int d^2\theta\,dt\,d^Dx\,\sqrt{G}N\left\{\alpha_RR^{(G)}+\alpha_\Phi G^{ij}\p_i\Phi\p_j\Phi+\alpha_\Lambda\right\}.
\label{eenewsw}
\ee
Any appearance of $T$ without derivatives would violate the shift symmetry, and spatial derivative terms are not available because $T$ is projectable.  Note that we have also required the absence of mixing terms between the $T$ sector and the $g_{ij}$, $n^i$ and $n$ sector of the theory.  Thus, the $T$ sector is trivial and decoupled from the metric geometry.  The localization equation for the lowest component field in $T$ simply states that this component field is constant in time.

Using the insights from effective field theory of inflation about the treatment of the Goldstone field associated with spontaneously broken time translations \cite{cheung,bmca,shiu}, we choose to define the component fields of $T$ as follows,
\be
T(t,\theta,\bar\theta)=t+\tau(t)+\theta\,\eta(t)+\bar\theta\,\bar\eta(t)+\theta\bar\theta\,b(t).
\ee
Note that we have defined the component field $\tau(t)$ such that the lowest component of the $T$ superfield is $t+\tau(t)$.  
The pair of real projectable fermions $\eta$, $\bar\eta$ are the ghost and the antighost, and $b$ is a projectable auxiliary field.  Since the BRST charge $Q$ acts on $\tau$ as $Q\tau=\eta$, this BRST multiplet clearly originates from an underlying projectable topological symmetry acting on $\tau$,
\be
\delta\tau(t)=\mathrm{f}(t),
\ee
with $\mathrm{f}(t)$ an arbitrary real function of $t$.  Together with the underlying gauge symmetries of our topological theory of the metric multiplets, the full gauge symmetry of the theory including the $\tau$ sector continues to be non-redundant, and $Q$ continues to play the role of the standard BRST charge, as in \cite{gst}.

With this choice of variables, the localization equation in the $T$ sector now gives
\be
\frac{\p}{\p t}\left[t+\tau(t)\right]=0,
\ee
which we will rewrite in the following suggestive form,
\be
\frac{\p}{\p t}\tau=-1.
\label{eetauflow}
\ee
This is indeed the last of Perelman's flow equations (\ref{eepfw}) for the shrinking solitons in the context of the $\CW$ functional.  

In the effective field theory of cosmological inflation \cite{cheung,bmca,shiu}, the Goldstone field that corresponds to our $\tau$ is traditionally called $\pi$; more importantly, in inflationary cosmology this field $\pi$ is nonprojectable.  It is interesting to note that in our construction of topological quantum gravity of the Ricci flow, it is also possible to promote $T$ to a nonprojectable field.  This would lead to two modifications of the action:  First, the kinetic term needs to be covariantized under spatial diffeomorphisms, and integrated over the entire spacetime:
\be
\int dt\,d^Dx\,d^2\theta\,\sqrt{G}\,\bar\CD T\,\CD T,
\label{eesknonpr}
\ee
with the covariant derivatives $\CD T$ and $\bar\CD T$ given by
\bea
\CD T=\RD T-S^i\p_iT,\\
\bar\CD T=\barRD T-\bar S^i\p_iT.
\eea
Secondly, when we still keep the constant shift symmetry (\ref{eecshift}), it is now possible to add new terms to the superpotential $S_\CW$,
\be
\int dt\,d^Dx\,d^2\theta\,\sqrt{G}\left\{\alpha_T\,\p_iT\p_iT+\ldots\right\}.
\label{eeswnonpr}
\ee
Here we have indicated just the simplest, lowest-derivative term quadratic in $T$ and consistent with the shift symmetry.  In principle, one should also consider the possibility of mixing between the $T$ sector and the metric sector, both in the kinetic and the superpotential terms.  The localization equation that corresponds to (\ref{eesknonpr}) and (\ref{eeswnonpr}) is 
\be
\nabla_t\tau=-1+\alpha_T\,\Delta\tau.
\label{eeheat}
\ee
Writing $\tau=-t+\ST(t,x^i)$, (\ref{eeheat}) becomes simply the covariant heat equation $\nabla_t\ST=\alpha_T\Delta\ST$ on $\Sigma$:  Under the spatial diffeomorphism of $\Sigma$, $\ST$ transforms as a scalar, and $\Delta$ is thus the Laplacian of $g_{ij}$ on scalars.  While such nonprojectable extensions of our theory are indeed possible, for the purposes of the present paper we see no advantage in extending the theory to nonprojectable $T$, and therefore we will consider only the case of projectable $T$ from now on.  

\subsection{Perelman's equations for shrinking solitons from topological gravity}

The transformation between the fields $g_{ij}$ and $\phi$ of topological gravity and Perelman's variables $\hat g_{ij}$ and $\hat\phi$ will now have to involve factors of $\tau$.  On the topological gravity side, our equations are the same as in Section~\ref{seccpf}, schematically of the form (\ref{eesch}).  No $\tau$ has been introduced yet -- the entire dependence on $\tau$ will come from rewriting the theory in Perelman's variables.  Therefore, in order to match the leading terms on the two sides of (\ref{eesch}), we must again set
\be
\hat g_{ij}=e^\phi g_{ij},
\ee
as we did in Section~\ref{secff}.  We need another relation to determine the change of variables uniquely.  The key is again to look at Perelman's fixed-volume condition, (\ref{eewvol}), which suggests that we identify 
\be
g_{ij}=\frac{1}{4\pi\tau}\hat g_{ij}e^{-2\hat\phi/D}
\ee
and interpret (\ref{eewvol}) as the condition of time independence of the spatial volume element $\sqrt{g}$ in topological gravity -- a condition we know how to ensure dynamically, by going to the $|\lambda|\to\infty$ limit.  

Putting these two conditions together, we obtain our transformation rules between the two sets of fields,
\bea
\hat g_{ij}&=&e^\phi g_{ij},\\
\hat\phi&=&\frac{D}{2}\left[\phi-\log(4\pi\tau)\right].
\eea
In addition, the shift vector and the $\tau$ field stay unchanged: $\hat n^i=n^i$, $\hat\tau=\tau$.  Note that in the inverse of this transformation, $g_{ij}$ depends explicitly on $\tau$:
\bea
g_{ij}&=&\frac{1}{4\pi\tau}e^{-2\hat\phi/D}\hat g_{ij},\label{eegtau}\\
n&=&\frac{1}{4\pi\tau}e^{-2\hat\phi/D},\label{eephitau}
\eea
with our lapse function as always given by $n\equiv e^{-\phi}$.

With the transformation properties determined, we are in the position to rewrite our localization equations (\ref{eeponeour}), (\ref{eeptwoour}) and (\ref{eetauflow}) of topological gravity in Perelman's variables.  A direct calculation yields
\bea
\hat\nabla_t\hat g_{ij}&=&-2\hat R_{ij}-2\hat\nabla_i\p_j\hat\phi,\nonumber\\
\hat\nabla_t\hat\phi&=&-\hat R-\hat\Delta\hat\phi+\frac{D}{2\tau},\\
\dot\tau&=&-1.\nonumber
\eea
These equations are indeed the covariantized version of the flow equations (\ref{eepfw}) that Perelman derived from his $\CW$ entropy functional!  They reduce to (\ref{eepfw}) in Perelman gauge $\hat n^i=0$.  Thus, the conclusion is the same as in the case of the $\CF$-functional:  By taking the $|\lambda|\to\infty$ limit of our topological quantum gravity -- augmented now by the decoupled Goldstone superfield $T$ -- we get a covariantized version of Perelman's equations (\ref{eepfw}) associated with the $\CW$ entropy functional and the shrinking solitons.

What remains to do is to perform the alternate gauge fixing of spatial diffeomorphisms and to go to Hamilton gauge, in order to establish the relation with Perelman's equations (\ref{eepfwdt}) after the DeTurck trick has been performed on them.  The gauge fixing condition turns out to be the same (in either set of variables) as in Section~\ref{secdtt}:  In Perelman's variables, Hamilton gauge is
\be
\hat n_i=\p_i\hat\phi,
\ee
while in our variables it can be equivalently written as 
\be
n_i=-\frac{D}{2}\p_in.
\ee
Note that these relations imply that
\be
n_i=\frac{1}{4\pi\tau}e^{-2\hat\phi/D}\p_i\hat\phi,
\ee
which together with (\ref{eegtau}) and (\ref{eephitau}) gives the full list of our ADM fields describing the spacetime geometry of topological quantum gravity in Hamilton gauge in terms of Perelman's geometric data.

\subsection{Expanding solitons and the $\CW_+$ entropy functional}

Shortly after Perelman's work, Feldman, Ilmanen and Ni \cite{finexp} modified the $\CW$ entropy functional to a form suitable for expanding Ricci solitons.  Their $\CW_+$ entropy functional depends on $\hat g_{ij}$, $\hat\phi$ and instead of $\tau(t)$ a new projectable field $\sigma(t)$ which will now be a \textit{growing} linear function of time.  It takes the form
\be
\CW_+=\int d^D x\sqrt{\hat g}\frac{1}{(4\pi\sigma)^{D/2}}e^{-\hat\phi}\left\{\sigma\left(\hat R+\hat g^{ij}\p_i\hat\phi\p_j\hat\phi\right)-\hat\phi +D\right\}.
\ee
The volume-fixing condition is to hold the following measure,
\be
\frac{1}{(4\pi\sigma)^{D/2}}e^{-\hat\phi}\sqrt{\hat g}\,d^D x,
\ee
fixed in time, and the corresponding gradient flow equations are
\bea
\frac{\p\hat g_{ij}}{\p t}&=&-2\hat R_{ij}-2\hat\nabla_i\p_j\hat\phi,\nonumber\\
\frac{\p\hat\phi}{\p t}&=&-\hat R-\hat\Delta\hat\phi-\frac{D}{2\sigma},
\label{eewpluse}\\
\frac{\p\sigma}{\p t}&=&+1.\nonumber
\eea
The last of these equations is solved by setting $\sigma=t-t_0$ for some constant $t_0$, and that is how $\sigma$ is often interpreted in the mathematical literature.  After DeTurck's trick, these equations are simplified to
\bea
\frac{\p\hat g_{ij}}{\p t}&=&-2\hat R_{ij},\nonumber\\
\frac{\p\hat\phi}{\p t}&=&-\hat R-\hat\Delta\hat\phi+\hat g^{ij}\p_i\hat\phi\p_j\hat\phi-\frac{D}{2\sigma},
\label{eetildet}\\
\frac{\p\sigma}{\p t}&=&+1.\nonumber
\eea

It is easy to see how this set of equations can be reproduced in our topological quantum gravity.  Using the same action (\ref{eenewsk}) and (\ref{eenewsw}) as in the case of shrinking solitons, we now define the components of the Goldstone superfield $T$ as
\be
T(t,\theta,\bar\theta)=t-\sigma(t)+\theta\,\eta(t)+\bar\theta\,\bar\eta(t)+\theta\bar\theta\,b(t).
\ee
In terms of these components, the localization equation for $\sigma$ reads
\be
\frac{\p\sigma}{\p t}=+1,
\label{eesigmaflow}
\ee
which reproduces the last equation in (\ref{eetildet}).  Next, we propose the following change of variables from our fields to those of the $\CW_+$ functional,
\bea
\hat g_{ij}&=&e^\phi g_{ij},\\
\hat\phi&=&\frac{D}{2}\left[\phi-\log(4\pi\sigma)\right],
\eea
again keeping the lapse $n^i$ and the projectable field $\sigma$ unchanged: $\hat n^i=n^i$, $\hat\sigma=\sigma$.  In these Perelman-like variables, our localization equations (\ref{eeponeour}), (\ref{eeptwoour}) and (\ref{eesigmaflow}) are found to be
\bea
\hat\nabla_t\hat g_{ij}&=&-2\hat R_{ij}-2\hat\nabla_i\p_j\hat\phi,\nonumber\\
\hat\nabla_t\hat\phi&=&-\hat R-\hat\Delta\hat\phi-\frac{D}{2\sigma},\\
\dot\sigma&=&+1.\nonumber
\eea
Thus, we again find the perfect match between the localization equations of topological quantum gravity, and the covariantized flow equations for the expanding solitons associated with the $\CW_+$ functional.  Going to Perelman or Hamilton gauge will again establish the isomorphism with the $\CW_+$ flow equations (\ref{eewpluse}) before or (\ref{eetildet}) after DeTurck's trick.

\section{Conclusions}

In this paper, we identified the precise regime of nonrelativistic topological quantum gravity of \cite{grf}, in which the localization equations in the path integral are identical to Perelman's celebrated Ricci flow equations.  This map involves an interesting change of frames between Perelman's geometric variables and ours.  Perelman's fixed-volume condition is implemented by taking the ``farpoint'' limit $\lambda\to-\infty$ in the kinetic coupling of our nonrelativistic topological gravity.

When the localization equations correspond to the Ricci flow equations, the quantum theory exhibits anisotropic scaling between time and space, characterized by dynamical exponent $z=2$, for any spacetime dimension $D+1$.  Such a theory would be power-counting renormalizable for $D=2$, but its mathematical structure is richer and more relevant to deep questions of topology and geometry when studied for $D=3$.  This raises an intriguing question of a short-distance completeness, and possibly renormalizability of this quantum field theory that goes beyond naive perturbative power counting.  In which dimensions is our topological quantum gravity UV complete?  Does its topological BRST symmetry play a role in improving the short-distance behavior of the path integral?  Such questions remain open for closer examination.

Perhaps the main importance of the precise embedding of Perelman's Ricci flow theory into topological nonrelativistic quantum gravity that we identified in the present paper stems from the fact that it sets the stage for importing the remarkable wealth of mathematical results accumulated in the theory of Perelman's Ricci flow over the past two decades into the physical setting of quantum gravity, at least in the relatively well-controlled situation of a topological theory with no local propagating degrees of freedom.  The structure of solutions of Perelman's equations is well-studied, and exhibits many fascinating dynamical features, including topology-changing processes and other deeply nonequilibrium phenomena, not only in the most analyzed case of $3+1$ spacetime dimensions.  The frequent appearance of various ``entropy functionals'' with precise monotonicity properties along the general flow is also begging for an explanation more directly grounded in the physical arguments of quantum gravity and quantum field theory.  For instance, in the present paper we found an intimate relation between the $\CW$ and $\CW_+$ entropy functionals associated with the mathematical theory of the shrinking and expanding Ricci solitons, and the important physical concept of spontaneous symmetry breaking.  We expect that future investigations will continue this process of mutual illumination between the physical and mathematical aspects of the theory.

\acknowledgments

We wish to thank William Linch III for many illuminating discussions during the course of this work.  The main results of this work were presented by P.H. at the \textit{Marty Halpern Memorial Symposium} at Berkeley, California, in March 2019; and at \textit{FQMT 19: Frontiers of Quantum and Mesoscopic Thermodynamics} in Prague, Czech Republic, in July 2019.  P.H. thanks the organizers for the invitation and for creating a stimulating environment, and the conference participants for useful discussions.  This work has been supported by NSF grants PHY-1820912 and PHY-1521446.  

\appendix
\section{Collection of the reframing formulas}
\label{ssapp}

The map between Perelman's metric $\hat g_{ij}$ and dilaton $\hat\phi$ on one hand, and our fields $g_{ij}$ and $\phi$ on the other is given by a nonlinear transformation of the fields, which involves a change of frame of the metric.  Such changes of frame are common in theories of gravity coupled to scalar fields, in particular in string theory.  It is well-known that different geometric probes (such as branes of diverse dimensions) may be naturally probing the spacetime geometry in distinct frames.

In the present context of nonrelativistic quantum gravity, the slight novelty of the change of frames stems from the fact that we are reframing the \textit{spatial} metric $g_{ij}$, and that the role of the spatial scalar field is played by the (logarithm of the) lapse function $n$.  Of course, the formulas for the transformation properties of various geometric objects under such a change of frame are standard and well-understood.  We collect them in this Appendix simply for convenience and completeness, given the prominent role that they play in the precise comparison between our theory and Perelman's equations in the bulk of the paper.

We begin with a spatial metric $g_{ij}$ on a $D$-dimensional manifold $\Sigma$, and a field $\phi$ which transforms under the spatial diffeomorphims of $\Sigma$ as a scalar.  A ``change of frame'' from $g_{ij}$ to $\tilde g_{ij}$ is defined as the transformation
\be
\tilde g_{ij}=e^{\alpha\phi}g_{ij},
\ee
for some real constant $\alpha$.  We will systematically denote all the geometric objects in the new frame $\tilde g_{ij}$ by $\tilde{\ \ }$.  The tilde and un-tilde quantities are related as follows.  The volume element is given by
\be
\sqrt{\tilde g}=e^{\alpha D\phi/2}\sqrt{g}.
\ee
The Christoffel symbols of the Levi-Civit\`a connections of $\tilde g_{ij}$ and $g_{ij}$ are related by
\be
\tilde\Gamma^k_{ij}=\Gamma^k_{ij}+\frac{\alpha}{2}\left(\delta^k_j\,\p_i\phi+\delta^k_i\,\p_j\phi-g_{ij}g^{k\ell}\p_\ell\phi\right).
\ee
The Riemann tensor is given by
\bea
\tilde R^i{}_{jk\ell}&=&R^i{}_{jk\ell}+\frac{\alpha}{2}\left(\delta^i_\ell\nabla_k\p_j\phi
-\delta^i_k\nabla_\ell\p_j\phi+g_{jk}g^{im}\nabla_\ell\p_m\phi-g_{j\ell}g^{im}\nabla_k\p_m\phi\right)\nonumber\\
&&\quad{}+\frac{\alpha^2}{4}\left(\delta^i_k\,\p_\ell\phi\,\p_j\phi-\delta^i_\ell\,\p_k\phi\,\p_j\phi
+\delta^i_\ell\, g_{jk}g^{ms}\p_m\phi\,\p_s\phi\right.\nonumber\\
&&\qquad\left.-\delta^i_k\, g_{j\ell}g^{ms}\,\p_m\phi\,\p_s\phi+g_{j\ell}g^{is}\,\p_s\phi\,\p_k\phi-g_{jk}g^{is}\,\p_s\phi\,\p_\ell\phi\right),
\eea
the Ricci tensor by
\bea
\tilde R_{ij}&=&R_{ij}+\frac{\alpha}{2}\left\{(2-D)\nabla_i\p_j\phi-g_{ij}\Delta\phi\right\}\nonumber\\
&&\quad{}+\frac{\alpha^2}{4}\left\{(D-2)\p_i\phi\,\p_j\phi+(2-D)g_{ij}(g^{k\ell}\p_k\phi\,\p_\ell\phi)\right\},
\eea
and the Ricci scalar is
\be
\tilde R=e^{-\alpha\phi}\left\{R+\alpha(1-D)\Delta\phi-\frac{\alpha^2(D-2)(D-1)}{4}g^{ij}\p_i\phi\,\p_j\phi\right\}.
\ee
The Laplace operator on scalars transforms as 
\be
\tilde\Delta f=e^{-\alpha\phi}\left\{\Delta f+\frac{\alpha(D-2)}{2}g^{ij}\p_i\phi\,\p_j f\right\}.
\ee
Finally, in the bulk of the paper we also need the relation between the uncontracted second derivatives on a scalar,
\be
\tilde\nabla_i\p_jf=\nabla_i\p_jf+\frac{\alpha}{2}g_{ij}(g^{k\ell}\p_k\phi\,\p_\ell f)-\frac{\alpha}{2}(\p_i\phi\,\p_jf+\p_j\phi\,\p_if).
\label{eeuncontr}
\ee
%
\bibliographystyle{JHEP}
\bibliography{grf}
\end{document}